# Carbon Paste Electrode Modified Poly-Glutamic Acid (PGA) with Molecularly Imprinted for Detection of Rhodamine B

**Henry Setiyanto[1]\*, Ferizal Ferizal[1], Vienna Saraswaty[2], Ria Sri Rahayu[1], Muhammad Ali Zulfikar[1]**

[1]Analytical Chemistry Research Group, Institut Teknologi Bandung, Indonesia
[2]Research Unit for Clean Technology, Indonesian Institute of Science, Indonesia

\*Corresponden Email: henry@chem.itb.ac.id

**Abstract.** Rhodamine B is a synthetic dye used for coloring textiles, paper and ceramics. In addition, Rhodamine B is also often used for coloring ingredients in food ingredients such as crackers, syrups, candy, cakes, and is often used for coloring lipsticks. The accumulation of Rhodamine B in the body can cause liver, kidney and lymph damage. In this study, a modified carbon paste (EPK) electrode with molecularly imprinted polymers (glutamic acid) was developed for the determination of rhodamine B using potentiometric techniques. The modification was carried out by electropolymerization of glutamic acid monomer from a solution containing 3.0 mM glutamic acid and 1.0 mM rhodamine B in a phosphate buffer pH 7 using a cyclic voltammetry technique of 15 cycles in the potential range between -0.2 V - 1.8 V with a scan rate of 100 mV / sec. Measurements with these electrodes have optimum performance at pH 4. The Nerst factor is 29.2 mV / decade. The measurement range is $10^{-5}$ M – $10^{-2}$ M. With the detection limit of Rhodamine B which can be measured by the electrode is $8.91 \times 10^{-6}$ M. This electrode has a fairly good value of accuracy and precision, and does not have a significant difference. when compared with the UV-Vis spectrophotometric method.

## 1. Introduction

The use of Rhodamine B as a food coloring in several countries has been banned because of its effects on human health. These compounds are often used for analytical reagents in the fields of mining, metals, pharmaceuticals, microbiology and photochemical analysis [1]. In addition, this dye is often used for crackers, syrups, candy, several kinds of pastries and also for cosmetic dyes such as lipstick [2], and ink [3]. The accumulation of rhodamine content can cause cancer, liver, kidney and lymphoma damage [4].

Until now, the determination of rhodamine B has been carried out by well-established methods such as high perfomance liquid chromatography [5], Liquid chromatography - mass spectroscopy [6], solid phase extraction [2], fluorometric-high performance liquid chromatography [7], UV-Vis spectrophotometry [8] and UV-Vis high performance liquid chromatography [9]. These methods have several shortcomings, including the need to make test sample preparation that is not simple, reagents that are not cheap, the limited ability of the test equipment to detect rhodamine B in very small concentrations, time consuming and less selective. One of the methods that can be used in the determination of rhodamine has the advantages of being easy in the preparation process, cheaper, small in size and allows for a very small detection limit, and is easy to automate, is electrochemistry (potentiometry).

Electroanalytical techniques such as voltammetry, potentiometry, amperometry and others are part of the electrochemistry method. This technique has been successful in explaining chemical reactivity [10], increasing both sensitivity and selectivity [11]-[13]. The technique developed to increase the selectivity and sensitivity is to form molecularly imprinted polymers (MIPs). In this study, a modification of the carbon paste electrode was carried out using MIPs for the determination of rhodamine B.

**2. Methodology**
2.1. Chemicals, Solution and Equipment.
The chemicals, sample solutions and equipment used in this study were the same as those used in previous studies [14].

2.2. Preparation of Reference Electrode, Carbon Paste Electrode (CPE) and Carbon Paste Electrode Modified MIPS (CPE-MIPs)
In this study we used the results of electrode preparation, both the reference electrode and the working electrode from the results of the research we had done previously [14].

2.3. Determining Linear Range and Nernst Factors.
The Nerst factor was determined from the measurement results of the rhodamine B standard solution using the EPK-MIP working electrode. The optimization results were made a curve between potential (mV) to log [rhodamine B] in order to obtain a line equation in the following equation: y = a + bx.
Where the slope of curve (b) is the Nernst factor, where y is the solution potential, x is the log value of the analyte concentration.

2.4. Determination of Detection Limits.
The detection limit value is obtained by determining the equation of the non-linear line on the potential curve (mV) against log [rhodamine B]. The two line equations then determine the intersection point. If the intersection of the two lines is extrapolated to the x-axis, a log of rhodamine B concentration is obtained from the limit of the electrode.

2.5. Test for precision and accuracy.
The precision or probability test was carried out by measuring the rhodamine B solution with a concentration of $10^{-5}$ M, $10^{-4}$ M and $10^{-3}$ M measured using the optimum EPK-MIP at the optimum pH for 3 times. Precision is determined by calculating the standard deviation (SD) and the coefficient of variation (CV). CV = SD / (average concentrations).
The accuracy test is obtained from the percent recovery value. To obtain% recovery, rhodamine B solutions with concentrations of $10^{-5}$ M, $10^{-4}$ M, $10^{-3}$ M, and $10^{-2}$ M were measured using an electrode at the optimum pH so that the potential of each solution was obtained. % recovery = (measured concentration) / (real concentration) x 100%.

2.6. Determination of the Selectivity Coefficient.
The electrode selectivity was calculated by measuring the potential of the interfering solution namely Na-Benzoate, sucrose, and MSG at a concentration of $10^{-3}$ M (the actual concentration of the interfering solution) using the optimized EPK-MIPs electrode.

$$k_{i,j}^{pot} = \frac{the\ measured\ concentration\ of\ the\ interfering\ solution}{the\ actual\ concentration\ of\ the\ interfering\ solution}$$

2.7. Analysis of Rhodamine B in Sample Solutions.
Measurement of the sample solution for rhodamine B was carried out with a concentration of $10^{-5}$ M. Then the potential measurement was carried out using MIPs-modified carbon paste electrodes at optimal pH. Rhodamine B analysis was also performed using UV-Visible spectrophotometry at a wavelength of 554 nm.

## 3. Result and Discussion
### 3.1. Measuring range
Whether an electrode is good or not can be determined from how wide the measurement range is. The wider it is, the better the electrode. Determination of the measurement range is carried out at the optimum EPK-MIP by comparing the Nernst factor and measurement range at different concentration ranges. The calculation results show a good measurement range of CPE-MIPs with a concentration of $10^{-5}$-$10^{-2}$ M with a Nernst factor of 29.2 mV / decade. Complete data are available in Table 1.

Table 1. CPE-MIPs Measurement Range

| Concentration (M) | Regression Line | Nerst Factor (Slope) | $R^2$ |
|---|---|---|---|
| $10^{-8}$ - $10^{-1}$ | y = 28.321x + 399.07 | 28.32 | 0.73 |
| $10^{-5}$ - $10^{-2}$ | y = 29.2x + 372.2 | 29.2 | 0.97 |
| $10^{-7}$ - $10^{-2}$ | y = 20,771x + 344,97 | 20.77 | 0.90 |

### 3.2. Detection Limit.
The detection limit is the smallest level of analyte in a sample that can still be measured by the instrument. Each selective electrode has the lowest and highest measurement limits which is the essential response of the electrodes. Determination of detection limits can be done by determining the point of extrapolation of the rhodamine B standard curve in the measurement range (Figure 1). In this study, the lower detection limit was obtained at $8.91 \times 10^{-6}$ M. This shows that the CPE-MIPs has excellent performance and is very sensitive. This electrode is capable of detecting very small concentrations.

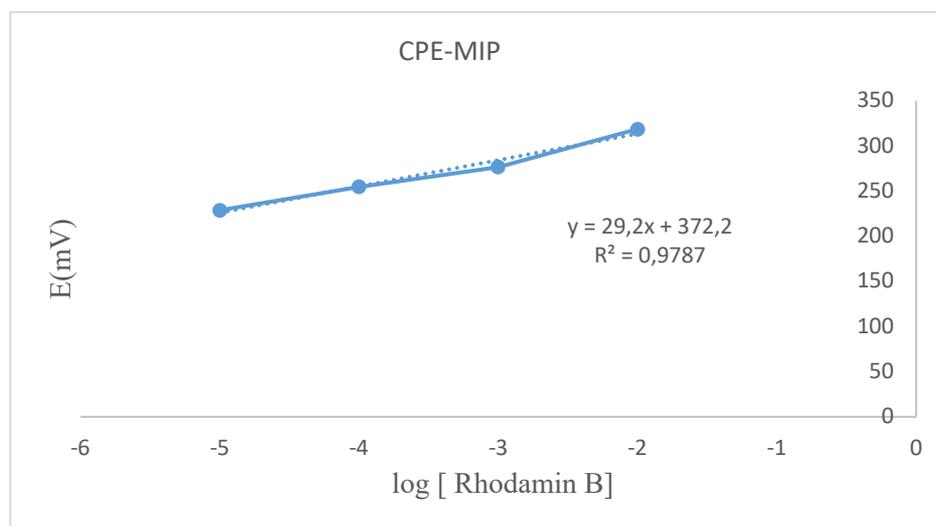

Figure 1. Calibration curve of Rhodamine B

3.3. Precision Test.
The work electrodes made in this study were tested for their reliability in measuring the standard solution of rhodamine B with concentrations of $10^{-5}$ M, $10^{-4}$ M, and $10^{-3}$ M and measured its potential three times. For the potentiometric method, the precision is said to be good if the CV value is between 1-3% [15]. The better the precision value is the lower the CV value. The full data is provided in Table 2.

**Table 2.** Coefficient of Variation of Measurement

| Concentration (M) | E (mV) | DS | CV (%) |
|---|---|---|---|
| $10^{-5}$ | 219.8 | 2.80 | 1.27 |
| $10^{-4}$ | 244.9 | 2.76 | 1.12 |
| $10^{-3}$ | 273.5 | 3.27 | 1.19 |

3.4. Determination of % recovery (% R).
The percentage of recovery shows the accuracy of a measurement. The measurement accuracy of Rhodamine B was determined using four concentrations, namely $10^{-5}$ M, $10^{-4}$ M, $10^{-3}$ M and $10^{-2}$ M. Potential data and measurement accuracy are shown in Table 3.

**Table 3.** Measurement Accuracy

| Concentration (M) | E (mV) | Recovery (%) |
|---|---|---|
| $10^{-5}$ | 225 | 91.2 |
| $10^{-4}$ | 255 | 97.2 |
| $10^{-3}$ | 283 | 86.7 |
| $10^{-2}$ | 312 | 87.0 |

The results show that EPK-MIPs has good accuracy for both concentrations because it is still in the tolerable range (90-107 %).

3.5. Electrode Selectivity.
Potentiometric electrodes have a selective character for certain analytes. The level of selectivity of an electrode is determined by the value of the selectivity coefficient. The coefficient of selectivity was determined by measuring the potential of interfering solutions such as Na-Benzoate, sucrose and MSG at a concentration of $10^{-3}$ M. The amount of the selectivity coefficient follows the following conditions. If the value of $k_{i,j} < 1$, then the electrode is selective towards i ion rather than j ion, and vice versa [16]. Table 4 showed that the carbon paste electrode MIPs (CPE-MIPs) had excellent selectivity. If the selectivity coefficient is less than $10^{-3}$ indicates that the electrode is very selective against rhodamine B, it is not selective against Na-Benzoate, sucrose and MSG.

**Table 4.** Selectivity coefficient measurement

| Compound | Potential (mV) | Concentration (M) | Measured Concentration (M) | $k_{i,j}$ |
|---|---|---|---|---|
| Na-Benzoat | 172.0 | 0.001 | 1.393 x $10^{-7}$ | 1.393 x $10^{-4}$ |
| Sucrose | 155.0 | 0.001 | 3.715 x $10^{-7}$ | 3.715 x$10^{-4}$ |
| MSG | 147.4 | 0.001 | 2.001 x $10^{-8}$ | 2.00 x $10^{-5}$ |

3.5. Rhodamine B analysis used the Potentiometric method and the UV-Visible Spectrophotometric method.

The recovery percentage in Table 5 provides information that the two methods are not much different. In addition, statistical tests can also be done to compare the two methods. The comparison of the variance of the two methods was determined by using the F-Test and T-Test.

**Table 5.** % Recovery for Measurement of $10^{-5}$ M Rhodamine B Solution

| Methods | Recovery (%) | | |
|---|---|---|---|
| | Sample 1 | Sample 2 | Sample 3 |
| Potentiometric | 83.83 | 99.40 | 108.2 |
| Spektrophotometry Uv-Vis | 95.13 | 99.97 | 99.00 |

From the calculation, the F-count value is 23.11 and the F-table at the 95% confidence level is 39.00 so that the T-count value can be determined. The T-table value at the 95% confidence level is 4.30 while the T-count value is 0.12. Because the value of T-table is greater than T-count, the null hypothesis is accepted. This means that there is no significant difference in the concentration of Rhodamine B obtained from the two methods.
.

**4. Conclusion**

Based on the results of this study, it can be concluded that MIPs modified carbon paste electrodes (CPE-MIPs) with rhodamine B as the template molecule can be made using the cyclic voltammetry method to analyze rhodamine B potentiometrically.

Rhodamin B analysis by the potentiometric method using a working electrode of carbon paste modified MIP has a measurement range of $10^{-5}$ M - $10^{-2}$ M, detection limit of $8.91 \times 10^{-6}$ M, has a fairly good value of accuracy and precision. These results do not have a significant difference with the UV-Vis spectrophotometer method.